# Developing a Trust Domain Taxonomy for Securely Sharing Information Among Others

Nalin Asanka Gamagedara Arachchilage, Cornelius Namiluko, Andrew Martin
Department of Computer Science
University of Oxford,
Wolfson Building, Parks Road, Oxford, UK
nalin.asanka@cs.ox.ac.uk, cornelius.namiluko@cs.ox.ac.uk, andrew.martin@cs.ox.ac.uk

*Abstract*—**In any given collaboration, information needs to flow from one participant to another. While participants may be interested in sharing information with one another, it is often necessary for them to establish the impact of sharing certain kinds of information. This is because certain information could have detrimental effects when it ends up in wrong hands. For this reason, any** *would-be participant* **in a collaboration may need to establish the guarantees that the collaboration provides, in terms of protecting sensitive information, before joining the collaboration as well as evaluating the impact of sharing a given piece of information with a given set of entities. The concept of a trust domains aims at managing trust-related issues in information sharing. It is essential for enabling efficient collaborations. Therefore, this research attempts to develop a taxonomy for trust domains with measurable trust characteristics, which provides security-enhanced, distributed containers for the next generation of composite electronic services for supporting collaboration and data exchange within and across multiple organisations. Then the developed taxonomy is applied to possible scenarios (e.g. Health Care Service Scenario and ConfiChair Scenario), in which the concept of trust domains could be useful.**

## I. INTRODUCTION

Why trust is so important in online activities? Trust is an element that entails in online exchange relationships characterized by uncertainty, anonymity, a lack of control and potential opportunism [1]. In online relationships, such as online transactions often require sharing of sensitive information (e.g. username, password, emailing address or telephone number), corporate information (e.g. airline reservation or inventory data) and financial information (e.g. credit or debit card numbers) among the transacting parties. Therefore, trust helps to ensure that one does not take an advantage of others during or after the particular transaction.

There has been an increasing interest in the imperative of trust in societies in general and in organizations in particular [2], [3], [4], [5], [6], [7], [8], [9], [10], [19], [20]. However, there is no common integrative taxonomy to interpret the extant literature and to guide future research regarding online trust related issues [4], [19], [20] . Li [11] offers a systematic review; other authors [12], [13] present a meta-analysis. Clearly, some aspects of trust are subjective; moreover there are many contextual factors in play. It is often related to and also affected by individuals' differences and their situational factors [9]. It is true that different people interpret the role of trust quite differently in various situations and have a different scale of trust towards different trustees. The *Trusted Computing Group* defines trust by saying "an entity can be trusted if it always behaves in the expected manner for the intended purpose"[14].

Information sharing is an essential part of any business process. Every day, organizations and individuals exchange information for the purpose of service delivery, communication and collaboration [21]. For example, a researcher wishing to set-up an experiment may write an email to an infrastructure provider, describing the experiment and specifying the resource requirements for the experiment. This experiment could be of a sensitive nature, such as development of a vaccine or stem cell research and therefore requires that its content, intention or existence be known only to a selected list of individuals and organizations. Similarly, two law enforcement agencies working on similar cases may require sharing information about the evidence of a crime. However, each agency may need to share such information with only a selected list of other agencies or individuals from these agencies. In this case, how does one on the selected email list ensure the information shared by someone else hasn't been altered or contaminated by intruders? In general, we tend to trust people on our email list [21] [22]. However, there might be a hidden risk of contaminating the information (e.g. email message) during online transmission. A popular maxim captures the essence: "Trust is good, but control is better" [15]. There is, therefore, a need for mechanisms that allow parties to share such information while restricting its flow to the selected entities.

The research work reported in this paper introduces an integrative taxonomy for trust domains. The concept of trust domains is meant to be used for proving a foundation (evidence) for securely sharing information (how, when and with whom) among a group of entities. This enables the parties involved and the observers to understand the level of trust before going ahead with sharing data.

The reminder of this paper is structured in the following manner. Section II describes the elements of trust domains. Section III discusses the useful infrastructure models to develop our proposed taxonomy for trust domains. In section IV, we introduce a taxonomy for trust domains, which enables to securely share the information among the others. Then the introduced taxonomy is applied to a possible scenario in Section V. Section VI provides conclusions and opens up opportunities for future work that may extend the research work reported in this paper.

## II. ELEMENTS OF TRUST DOMAINS

In this paper, we are interested in providing these mechanisms in a way that enables the parties involved in information sharing to understand: i) the level of security that will be achieved together with the risks mitigated; ii) the mechanisms used to achieve it; and iii) economic, performance and scalability implications. Most importantly, these mechanisms should allow these parties to evaluate the above characteristics before they go ahead with sharing the information. To achieve this, we propose the notion of a *trust domain* as a concept that makes this possible - enabling organizations and individuals to maximize their business profitability through efficient and secure information sharing.

To start, we need to identify and define concepts that could help us capture the notion of a trust domain. In this paper, we are interested in identifying and characterizing these concepts in order to: i) provide perspectives from which to discuss the concept of a trust domain; ii) enable instantiation of trust domains with varying properties; and iii) provide integration and interoperability mechanisms. We build a model that captures the concepts that could be used to describe a trust domain and relationships among those concepts.

The model assumes that a trust domain will require some social structure as well as infrastructure support. The social structure provides regulatory and monitoring mechanisms for individuals and organizations involved in a trust domain as well as a means for assigning responsibility and appropriate penalty in the event that information flows to unwanted parties. The infrastructure, on the other hand, serves to provide mechanisms that ensure that technical systems involved in or used as a means of processing, communicating or sharing information observe and enforce the required information flow constraints. Therefore, we take an approach in which we identify some of the existing mechanisms that could serve as fundamental building blocks for a trust domain and then we find possible integration points to enable the conceptualization of a trust domain.

## III. USEFUL INFRASTRUCTURE MODELS

Our taxonomy for trust domains embraces systems deployed on diverse infrastructures. Each infrastructure may employ different technologies to serve the needs of a trust domain. However, at the architectural level, several aspects can be identified to be common among the various types of infrastructure. Of interest to develop taxonomy for trust domains are the Web Services Architecture [16] and the Trusted Multi-tenant Infrastructure (TMI) [17].

### A. Web services Architectural Model

The Web Services (WS) Architecture [16] defines functional components, relations among them and constraints upon which they operate. The architecture defines four models (message-oriented, policy, service-oriented and resource) that capture concepts and relations among them as viewed from a certain perspective. We collect the concepts defined in these models into an integrated architectural model as illustrated in Figure 1.

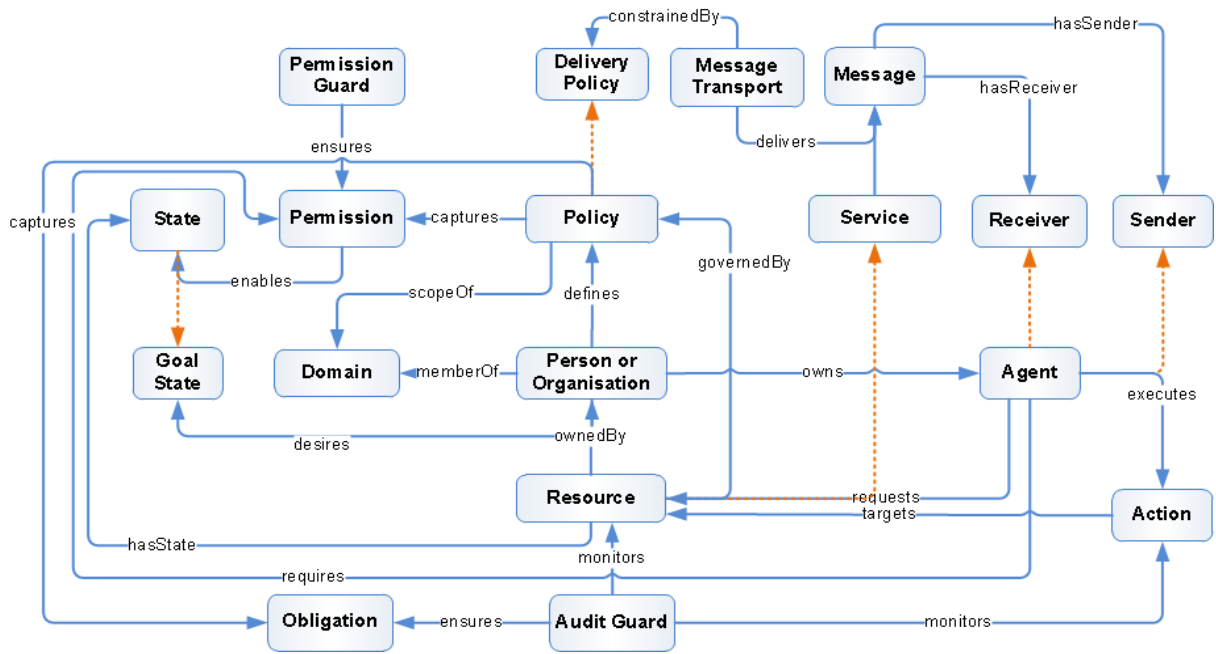

Figure 1: Integrated model for the WS Architecture - showing the link between service, message, resource and policy models

The message-oriented model defines *Message* as a first class concept that has a *Sender* and one or more *Receivers*. A message is delivered by some *MessageTransport* mechanism, which is constrained by a *DeliveryPolicy*. The *Delivery Policy* is a subclass of *Policy*, defined in the policy model, which defines *MessageReliability* properties of the *MessageTransport* mechanism. *Messages* are used as the main mechanisms for communicating requests for services and responses to the requests - where a service is defined, in the service-oriented model, as an abstract resource capable of performing one or more tasks. This definition of a service allows us to link a service to a *Resource* defined in the resource model through a subclass relation.

A resource has *State* and is owned by a person or organization that defines a *Policy* that governs how a *Resource* is used or behaves. This policy is also defines the scope of a domain and is defined in a way that captures the desired goal state. *Policies* capture permissions that are required by certain agents in order to enable certain resources attain some state and obligations that agents have on the resources. *Permissions* are enforced by a *PermissionGuard* while *Obligations* are enforced by an *Audit Guard*.

B. Trusted System Domain Model

The TCG TMI Working Group [17] aims to build a set of specifications for trustworthy operation in multi-tenant infrastructures. As part of this effort, the working group has released the first version of the use cases for a TMI. We use these use cases to build a model that includes concepts and relations among the concepts defined in the use cases. The concepts capture the parts of a TMI while the relations capture the dependencies, causalities and interactions among the concepts necessary to capture the functionality of a TMI as described in the use cases.

The model, illustrated in Figure 2, is built around three main entities, *Policy*, *Asset* and *Role*. An *Asset* is an entity that is of value to a person or organization that participates in a TMI. Its value is defined by the entity that has an interest in it. Implying that the same asset could have different values depending on whose point of view this value is being computed. A *Policy* is a piece of data that defines the constraints on how the assets are used or accessed by entities within a TMI as well as the expected behavior of such entities. A *Role* is a set of responsibilities assumed by an entity in the system. It also defines the activities that any entity with such a role can perform.

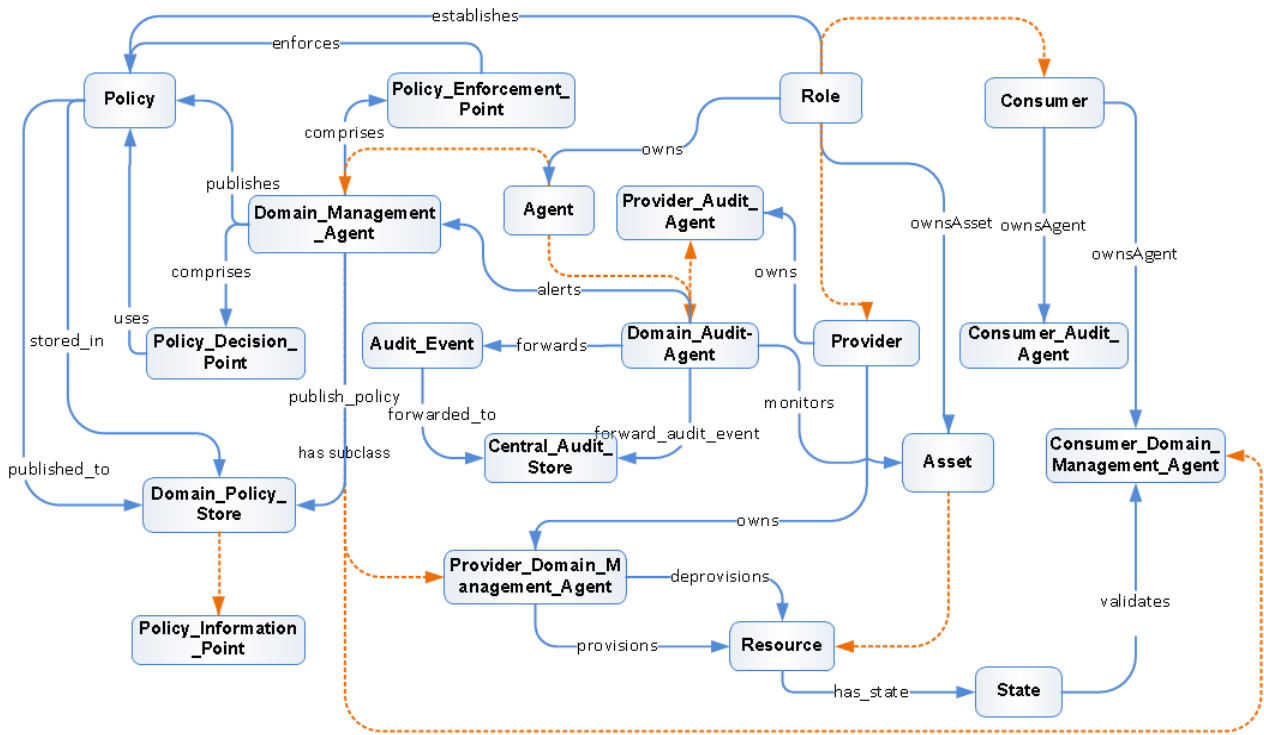

Figure 2: A model for the trusted multitenant infrastructure

TMI distinguishes between two main roles, *Consumer* and *Provider*, and defines several concepts some of which are generic across domains of the two roles while others are specific to each domain. A *DomainAuditAgent* is defined as a generic agent that forwards *AuditEvent* to the *CentralAuditStore* – a component that stores audit events collected from various parts of a TMI. The *ConsumerAuditAgent* and *ProviderAuditAgent* are subclasses of *DomainAuditAgent* that are specific to *Consumer* and *Provider* domains respectively.

Each *Role* establishes a *Policy* that is published to a *DomainPolicyStore* and also creates a *DomainManagementAgent* to manage the domain according to the policy they define. This *Policy* is expected to be enforced by a *PolicyEnforcementPoint* in the domain in which the policy is setup, and is monitored by the specific type of *DomainAuditAgent* for the particular domain. *Consumer DomainManagementAgent* and *ProviderDomainManagement Agent* are subclasses of *DomainManagementAgent* that are specific to *Consumer* and *Provider* domains respectively.

A *Resource* is defined as a subclass of *Asset*, which has *State* and can be provisioned or de-provisioned in the *Provider* domain by the *ProviderDomainManagementAgent*. To ensure that the correct policy is being enforced on a resource, the *Consumer* can validate the state of the resource using the *ConsumerDomainManagementAgent*.

IV. PROPOSED TRUST DOMAINS TAXONOMY

In this section of the paper, we discuss how the models identified above are combined to create trust domain taxonomy. We illustrate the concepts that can be used to integrate the models and discuss how the semantic gap in the usage of these concepts can be bridged.

*A. Fundamental Concepts and Relations*

The proposed model consists of a number of concepts, such that each concept captures a class of things that may exists in a trust domain, be used to build a trust domain and used within a trust domain. Though all these concepts may be used in different instances of a trust domain, a few of them can be identified as being fundamental to the existence of a trust domain. We identify the *Data, Policy, Controls, Roles, Actions* and *Evidence* as being fundamental concepts necessary to build a trust domain.

As depicted in Figure 3, a *Role* owns *Data* that will exist within a trust domain and establishes a *Policy* that constrains *Actions*.

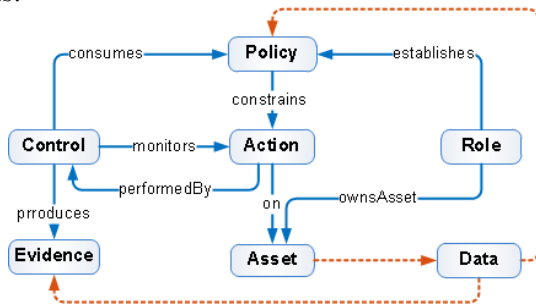

Figure 3: Fundamental concepts in a trust domain

As mentioned in the fundamental model, a *Role* establishes one or more policies within the domain. However, any given policy can only be established by one role. This means that if two roles establish identical policies, then both policies are treated as a unique entity, which can be linked through the equivalence property.

*Actions* are performed by a given role or by some agent that represents a particular role. These actions are monitored by *Controls* to ensure that the policy is being upheld. These controls produce *Evidence* to indicate that actions have been performed in accordance to the policies. Both *Evidence* and *Policy* can be considered to be a form of data, which can be manipulated in the same way as other data and may be subject to the same information flow restrictions. We describe each of these concepts in the following subsections.

*1) Asset*

Our conceptualization of a trust domain is based on the idea of enabling secure information flow among a set of entities. Such entities may each have a set of devices through which they share the data. Furthermore, these entities may provide access to the information stored on the devices or other media to other members of the domain. For this reason, we define the concept of an Asset as being a fundamental element of a trust domain. An asset is something of value to the owner, but could also be valuable to other entities such as attackers or competitors. One example of an asset is data. Data may exist in many different forms and indeed each form is faced with various kinds of challenges, with respect to controlling how it flows. For example, printed material could be prevented from being taken out of the building. However, it is still possible that someone could scan the materials and send it over a network or copy it to USB stick. For this reason, different types of protections maybe required to overcome each type of challenge. To reduce the scope of our work, we limit the model to data that exists in digital form. Other examples of assets include: resources such as computers; services such as web services; and communication infrastructure such as networks.

*2) Policy*

Policies are a means of specifying the behavior of entities within a trust domain and how data flows within or outside of a trust domain. Each policy specifies the expected relationship/dependency between one or more entities within a domain. For example, a policy can be used to specify the data allowed to into or out of a trust domain, the characteristics of the controls that should exist in a trust domain and the kind of evidence that these controls should provide. To satisfy the requirements for a trust domain, the policy being enforced, how it is being enforced and the decisions made in relation to this policy must be made explicit.

*3) Controls*

Controls are a set of mechanisms, processes or procedures that enforce the policies within a trust domain. These controls could be accomplished through social, e.g. penalties, or technical means e.g. trusted computing. Controls monitor activities that occur within a trust domain and produce evidence, described below, that can be used to determine the properties of a trust domain or its constituents.

*4) Roles*

Roles are used to specify the level of participation in a trust domain. Each role defines the kinds of activities that an entity assuming that role can do as well as the kinds of things that entities would be accountable for. Roles are also a way of separating concerns in a trust domain and isolating the activities that participants can perform in a trust domain.

*5) Actions*

Actions are a series of functionalities performed by components and agents in a trust domain. An action typically either consumes or produces some data or results in the change of state of a particular component or system. Depending on the level of abstraction desired, an action can be divided into sub-actions where each sub-action contributes to the overall outcome of the parent action.

*6) Evidence*

Evidence is data that is produced by the controls within a trust domain to indicate the kinds of activities that have occurred in a trust domain. These activities are captured by monitoring the actions that are performed by or on behalf of roles that exists within a trust domain. Examples of such evidence include; provenance - records of how data came to be; audit logs - log of events that occur during the lifetime of a trust domain; integrity measurement lists - a record of the binary hashes of software components and data in a trust domain; and digital certificates - specify the cryptographic identities of components that perform certain actions.

### B. Integrating Infrastructure Models

The fundamental concepts discussed above exist in more than infrastructure model. To build a taxonomy for trust domains that makes use of these models, we need to understand how the models can be integrated together. To achieve this, we use the fundamental concepts as integration points, so that each concept that exists in more than one model is linked using equivalence. This allows all the relations that apply to a concept in a particular model to apply to an equivalent concept in another model. As an example, *Policy*, which is defined in the fundamental model, exists in the trusted system domain model as well as in the policy model of the Web Service (WS)-architecture. This allows us to relate the policy concept in the Trusted System Domain (TSD) model to the *Permission* concept in the WS policy model, through the equivalence property between the two policy concepts and the captures property.

Using this approach, however, is faced with the challenge of a possibility of conflicts and inconsistencies. For example, in the WS policy model, the policy is defined by a person or organization while the TSD model specifies that a policy be established by a Role. To address this issue, we first identify that is a possible relationship between person or organization and role. Using this relationship, we can then determine which the appropriate concept to be linked directly to the common concept, i.e. Policy. In this particular case, we decided on Role because it is possible that other entities such as autonomous systems might be able to set-up a policy.

### C. Integrated Model

In this section of the paper, we describe an integrated model to form a taxonomy for trust domains resulting from the integration effort described above, also illustrated in Figure 4. We define a *DomainEntity* as a first class entity in a trust domain. It is defined as an entity that has a *memberOf* relation to the *Domain* and has exactly one of the following types: *Person, Organization, System, Process, Resource* or *Agent*. Furthermore, each *Domain EntityhasRole* within the domain. In this model, we do not restrict the membership and the role relationship, so that a domain entity may participate in more than one domain at any given time and may assume more than one role.

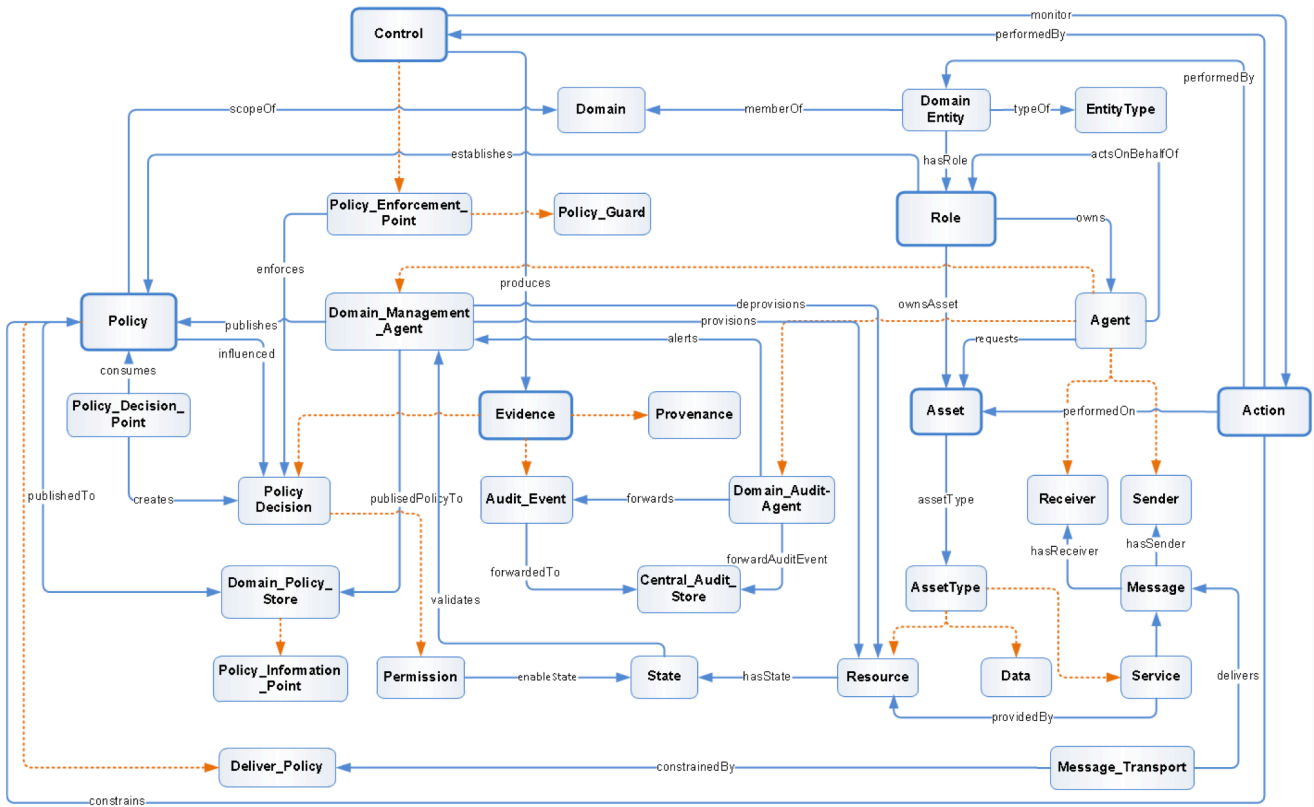

Figure 4: A taxonomy for trust domains

```
memberOf : DomainEntity ↔ Domain
hasRole : DomainEntity ↔ Role
typeOf : DomainEntity → EntityType
```

We define a *Policy* as data whose scope is limited to a *Domain*. In other words, a policy is only effective within the domain. It does not directly in the behavior or properties of entities outside the domain. This, however, raises a question of how to define identical domains (i.e. domains that enforce exactly the same kinds of policies). To answer this question, we define a *Policy* to be singleton object but allow it to be cloned; for example, copies can be made for use in other domains. The cloned policy can be linked to the original policy through the equivalence property.

```
scopeOf : Policy → Domain
enforces : PolicyEnforcementPoint ↔ PolicyDecision
subclass : PolicyEnforcementPoint → Control
subclass : Policy → Data
consumes : PolicyDecisionPoint ↔ Policy
creates : PolicyDecisionPoint ↔ PolicyDecision
publishes : DomainManagementAgent ↔ Policy
publishedTo : Policy → DomainPolicyStore
influenced : Policy ↔ PolicyDecision
publishedPolicyTo : DomainManagementAgent ↔ DomainPolicyStore

∀ p : Policy;  d1, d2 : DomainManagementAgent •
    (d1, p) ∈ publishes ∧ (d2, p) ∈ publishes ⇔ d2 = d1
∀ m : DomainManagementAgent;  s : DomainPolicyStore •
    (m, s) ∈ publishedPolicyTo ⇔
        (∃ p : Policy • publishedTo p = s ∧ (m, p) ∈ publishes)
∀ p : Policy;  x : PolicyDecision •
    (p, x) ∈ influenced ⇒
        (∃ y : PolicyDecisionPoint • (y, p) ∈ consumes ∧ (y, x) ∈ creates)
```

A policy is *publishedTo* the *DomainPolicyStore* by a *DomainManagementAgent* and is consumed by the *PolicyDecisionPoint*, which creates *PolicyDecisions*. *Policy Decisions* exist in the form of *Permissions* or *Obligations* and are enforced by a type of control called *PolicyEnforcementPoint*. The policy decisions can be considered as a type of part of evidence, which together with some policy meta-data can be used to support how the policy has been broken down into enforceable constraints.

An important aspect in trust domains is the ability to relate the decisions to the policies that triggered these decisions. For example, when a constraint specifies that a certain action be permitted, trust domains must be able to demonstrate how that constraint (i.e. to permit an action) was reached and the policies, which influenced this decision. For this reason, we define influenced as a relation between the set of policies and the set of decisions, which were used to arrive at the decisions. For this to work, we require a policy to be consumed by a *PolicyDecisionPoint* that creates the *PolicyDecision*.

```
owns : Role ↔ Agent
actsOnBehalf : Agent → Role
assetType : Asset → AssetType
performedOn : Action ⇸ Asset
performedBy : Action ⇸ DomainEntity ∪ Control
monitor : Control → Action

owns~ ∈ (Agent → Role)
```

A *Role* may also own *Agents*, which may act on their behalf, and *Assets*, which are shared with other members of the domain. *Assets* have a type, i.e. *AssetType,* which could either be *Resource*, *Data* or *Service*. Furthermore, the model specifies that individual assets and agents must be owned by exactly one *Role*.

*Actions* are performed on *Assets* by *Domain Entities* and controls within the domain. However, before an action can be performed, certain constraints determined by the policy must be satisfied. For this reason, we specify that *Policy constrainsAction*. To enable checking that appropriate constraints are satisfied before an action is performed, controls within the domain *monitor* actions.

```
validates : DomainManagementAgent ↔ State
hasState : Resource ↔ State
enableState : Permission ↔ State
provisions : DomainManagementAgent ↔ Resource
provisionedBy : Resource ⇸ DomainManagementAgent
deprovisions : DomainManagementAgent → Resource
subclass : AuditEvent → Evidence
forwards : DomainAuditComponent ↔ AuditEvent
alerts : DomainAuditComponent → DomainManagementAgent
forwardedTo : AuditEvent → CentralAuditStore
forwardsAuditEventTo : DomainAuditComponent → CentralAuditStore
```

*Resources* are an important aspect of trust domains. In this model, we define a resource as an asset type, which has state. A resource can be provisioned or de-provisioned by controls within the trust domain. When a resource is provisioned, it becomes available for use by agents and domain entities that request for *Assets* (including *Resources*) in the domain. The state of the resource also plays an important role in determining the properties of a trust domain. We therefore allow state to be validated by *DomainManagementAgent*. This validation determined whether or not the behavior or state of resources is in-line with domain policies.

A special type of control, referred to as *DomainAuditAgent,* is responsible for generating audit events by monitoring the activities in the domain. The *DomainAuditAgent* forwards *AuditEvent* to a *CentralAuditStore* where they can be analyzed as part of evidence to determine the properties of a trust domain and alerts the *DomainManagementAgent* when certain critical events are observed.

```
hasReceiver : Message ↔ Receiver
hasSender : Message → Sender
providedBy : Service ↔ Resource
subclass : Service ↔ Resource
delivers : MessageTransport ↔ Message
constrainedBy : MessageTransport ↔ DeliveryPolicy
subclass : DeliveryPolicy → Policy
```

We define messages as the main mechanisms for accessing services provided within the domain. Each message has a *Sender* and one or more *Receivers*; both defined as agents and are delivered through some *MessageTransport* mechanisms such as Remote Procedure Call (RPC) or Web Services, which is constrained by a *DeliveryPolicy*. The model defines a service as a *Resource*, which may be *providedBy* another *Resource*.

## V. APPLICATION SCENARIOS

In this section of the paper, we attempt to define possible scenarios such as Health Care Scenario and ConfiChair Scenario. Then the developed taxonomy is applied to possible scenarios, in which the concept of trust domains could be useful.

### A. Health Care Service Scenario

We consider a scenario, in which, a number of health institutions (i.e. hospitals or clinics) provide specialist services to their clients. Each service uses one or more databases to store data collected from interactions with their clients. The data collected maybe shared with other services, restricted to a particular service or isolated from other data stores within a single service. The specialist service providers are monitored for quality and financial purposes by Monitoring Services. For this reason, monitoring services may require access to certain parts of the data collected at a particular site. Furthermore, various monitoring services may need to share part of the data and keep some of it private.

We illustrate, in Figure 5, an example setup in which three specialist services share some parts of their data among themselves and two monitoring services that may access all or part of the data collected at each site. Each specialist service has a Demographics database and a number of other databases. In addition to the Demographics database, each service has the following other databases: Specialist Service 1 (SS1) has a Diagnostics database; Specialist Service 2 (SS2) has Scans and Treatment databases; and Specialist Service 3 (SS3) has Births database. Specialist Service 1 (SS1) and Specialist Service 2 (SS2) are monitored by Monitoring Service 1 (MS1) while Specialist Service 3 (SS3) is monitored by Monitoring Service 2 (MS2). Monitoring Service 1 (MS1) and Monitoring Service 2 (MS2), each have some databases where the monitoring data is stored.

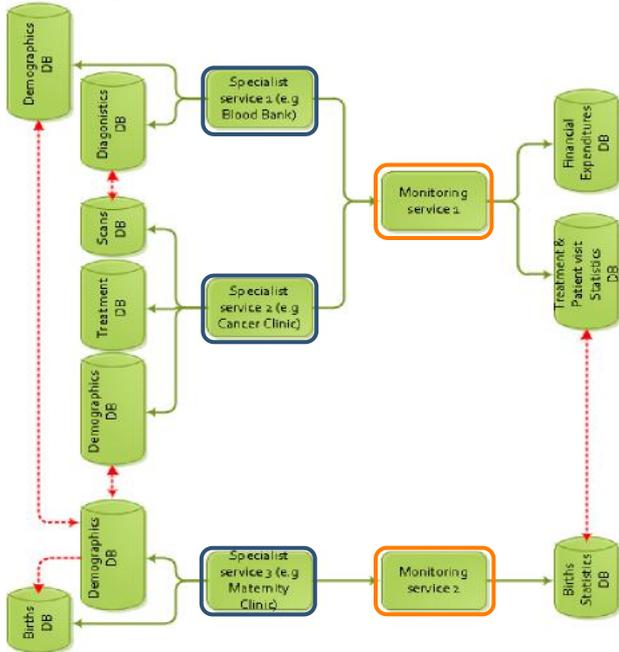

Figure 5: Information flows within health provision and monitoring

### 1. Data Flow Relations & Trust Domains Application

In this section of the paper, we attempt to apply the developed taxonomy for trust domains to the above-mentioned scenario (in this case Health Care Service Scenario). This is achieved by identifying the boundaries that exist with respect to data flows within the setup.

SS1 and SS3 have agreed on a policy to allow data to flow between their Demographics databases. SS3 has, additionally, agreed on a similar policy but with SS2. These results in two intersecting trust domains, SS1-SS3-Demo-TDom and SS2-SS3-Demo-TDom, are shown in Figure 6. According to the above-scenario, patients' demographics information (in these case "Assets") can only be shared between the blood bank and the cancer clinic. The maternity clinic has agreed on a similar policy but with only the cancer clinic. Therefore, this can be accomplished by defining policies, specifying the behaviour of entities within the trust domain and how data flows (in this case bi-directional) within the trust domain. Penalties or technical means (e.g. email warnings messages or systems generated warning messages) can be used to control activities that occur within the trust domain. These activities captured by monitoring the actions, can be used to produce evidence such as records of how demographics data came to be, audit logs-log of events that occur during the lifetime of a trust domain. Therefore, these controls produce evidence to ensure that actions have been performed according the policies defined as represented in the trust domains taxonomy.

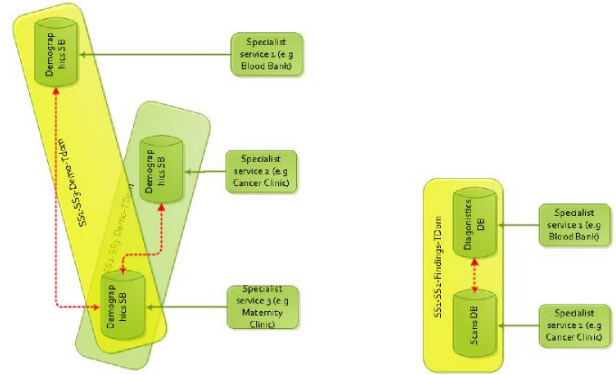

Figure 6: Trust Domains involving demographics data flow

SS1 and SS2 agree to share diagnostic and scan information between them. This agreement defines a bi-direction data flow between the two databases resulting in a trust domain, SS1-SS2-Findings-TDom, as shown in Figure 6. According to the above-scenario, patients' diagnostic and scan information (in this case "Assets") can only be shared between the blood bank and the cancer clinic. Therefore, this can also be accomplished by defining policies, specifying the behaviour of entities within the trust domain and how data flows (in this case bi-directional) within the trust domain. Penalties or technical means (e.g. email warnings messages or systems generated warning messages) can be used to control activities that occur within the trust domain. These activities captured by monitoring the actions can be used to produce evidence such as records of how diagnostic and scan information came to be, audit logs- log of events that occur during the lifetime of a trust domain. Therefore, these controls produce evidence to ensure that actions have been performed according to the policies defined as represented in the proposed trust domains taxonomy.

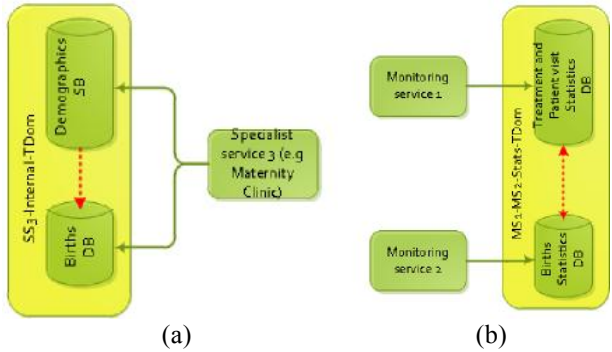

(a)    (b)

Figure 7: Internal SS3 Trust Domain (a), Monitoring statistics Trust Domain (b)

SS3 has an internal one-directional data flow from the Demographics database to the Births database. Data is not allowed to flow in the other direction. This creates a trust domain, SS3-Internal-TDom, as shown in Figure 7(a). According to the above-scenario, patients' demographics and birth information (in these case "Assets") can only be accessible by the maternity clinic. Additionally, the data is only allowed to flow from the Demographic database to the Birth database. Therefore, this can be accomplished by defining policies, specifying the behaviour of entities within the trust domain and how data flows (in this case one-directional) within the trust domain. Penalties or technical means (e.g. email warnings messages or systems generated warning messages) can be used to control activities that occur within the trust domain. These activities captured by monitoring the actions that can be used to produce evidence such as records of how demographics and birth information came to be, audit logs- log of events that occur during the lifetime of a trust domain. Therefore, these controls produce evidence to ensure that actions have been performed according the policies defined as represented in the proposed trust domains taxonomy. Finally, MS1 and MS2 have a mutual agreement to allow data to flow between their statistics databases, resulting in a trust domain, MS1-MS2-Stats-TDom illustrated in Figure 7 (b). According to the above-scenario, patients' birth as well as treatment and their visit statistics information (in this case "Assets") can only be shared between the two monitoring databases. Therefore, this can be accomplished by defining policies, specifying the behaviour of entities within the trust domain and how data flows (in this case bi-directional) within the trust domain. Penalties or technical means (e.g. email warnings messages or systems generated warning messages) can be used to control activities that occur within the trust domain. These activities captured by monitoring the actions that can be used to produce evidence such as records of how patients' birth as well as their treatment and visit statistics information came to be, audit logs- log of events that occur during the lifetime of a trust domain. Therefore, these controls produce evidence to ensure that actions have been performed according the policies defined as represented in the proposed trust domains taxonomy.

This case illustrates that trust domains will differ in their nature. Some will exist within a single entity such as an organisation or department while some will cross-organisational boundaries. Furthermore, some trust domains will place restrictions on the direction of data flow and others (e.g. the Demographics trust domains) will have some level of transitivity in their data flows. One thing common to all the domains, however, is that there is an agreement between the parties involved, which defines what data can flow within the domain as well as the direction of its flow. Other properties will definitely exist and this warranties the need for further investigations into the nature and expected properties of trust domains.

### B. *ConfiChair Scenario*

ConfiChair is a cloud-based conference management protocol developed by researchers from Birmingham University [18]. It is proposed to address a set of privacy and confidentiality requirements for existing cloud-based conference management systems such as EasyChair or the Editor's Assistant (EDAS). In ConfiChair, as shown in Figure 8, authors, reviewers and the conference chair interact through their browsers with the cloud, to perform the usual tasks of uploading and downloading papers and reviews. It is noteworthy that privacy and the data confidentiality issues concern the cloud-based conference system administrator who administrates the system for all conferences (not the conference chair, who is concerned with a single conference). For example, the cloud based conference system administrator could accidentally or deliberately disclose some sensitive information (e.g. strongest and weakest reviewers) because the service provider has access to all the information related to conferences hosted by the conference management system.

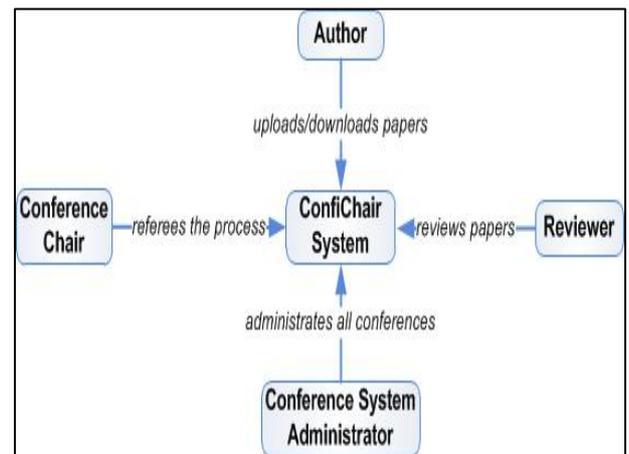

Figure 8: Trust Domains involving demographics data flow

*1. Data Flow Relations & Trust Domains Application*

In this section of the paper, we attempt to apply the developed taxonomy for trust domains to the above-mentioned scenario (in this case ConfiChair Scenario). This is achieved by identifying the boundaries that exist with respect to data flows within the setup.

According to the ConfiChair scenario shown in Figure 8, authors, reviewers, and the conference chair interact through their browsers with the cloud, to perform the usual tasks of uploading and downloading papers and reviews. Additionally, the conference system administrator is responsible of administrating the ConfiChair system for all conferences. Therefore, the *role* element in the taxonomy is represented in the ConfiChair scenario. Various roles carry some actions developed by the ConfiChair system. Author upload/download papers, reviewer reviews papers, conference chair referees the process and the conference system administrator administrates all conferences are a few examples. Therefore, the *action* element in the taxonomy is represented in the ConfiChari scenario. A set of policies developed within the ConfiChair system. For example, (a) a reviewer doesn't see other reviewers of a paper before writing her own, (b) the Conference System Administrator does not have access to the content of papers or reviews, or the numerical scores give by the reviewers to papers and (c) the conference system administrator does have access to the names of authors and the names of reviewers; however, he does not have ability to tell if a particular author was reviewed by a particular reviewer. The *policy* element in the taxonomy is, therefore, represented in the ConfiChair scenario. The following controls also developed in the ConfiChair system in order to monitor actions: (a) the login procedure implemented relies on each user having an identity *id* and a secret password *pswid;* (b)       secrecy of papers, reviews and scores: conference system administrator does not have access to the content of papers or reviews, or the numerical scores given by reviewers to papers; (c)       unlinkability of author-reviewer: conference system administrator does not have access to the names of authors and the names of reviewers and (d) however,     all sensitive data (e.g. encryption of each review or score) is seen by the conference system administrator only in encrypted form. Therefore, the *control* element in the taxonomy is also represented in the ConfiChair scenario. The following evidence such as information produced by uploading/downloading papers, reviewing papers, refereeing the ConfiChair system process and also administrating all conferences through the ConfiChair system is maintained by the ConfiChair system. The *evidence* element in the taxonomy is therefore represented in the ConfiChari scenario. Finally, data such as authors, reviewers, conference chairs and conference system administrator information; paper upload/download/submission information; paper review information; and referee process information exist in digital form within the ConfiChair system. Therefore, the *asset* element in the taxonomy is represented in the ConfiChair scenario.

The author uploads/downloads (in this case bi-directional) papers and the reviewers reviews (in this case bi-directional) those uploaded papers using the ConfiChair system. Therefore, this creates a trust domain, Author-ConfiChairSystem-Reviewer. According to the ConfiChair scenario shown in Figure 8, the conference chair referees the process. This creates two trust domains, Author-ConfiChairSystem-ConferenceChair as well as ConferenceChair-ConfiChairSystem-Reviewer.

Finally, the conference system administrator is responsible of administrating the ConfiChair system for all conferences where the conference chair referees the conference process. The conference system administrator does not have access to the names of authors and the names of reviewers and as well as all sensitive data (e.g. encryption of each review or score) is seen by the conference system administrator only in encrypted form. Therefore, this creates a trust domain, ConferenceChair-ConfiChairSystem-ConferenceSystemAdministrator.

## VI. CONCLUSIONS AND FUTURE WORK

This paper has presented a taxonomy for Trust Domains which enables individuals and organizations to securely collaborate across functions, geographies and corporate boundaries by providing collaborating parties (or participants) the means to create online environments designed to prevent information from leaking and where their resources can be shared as they specify. For this, Trust Domains relay on mechanisms ensuring that information flow to the right entities and that information is only accessed by entities capable of protecting it as specified. Furthermore, for future research, to help participants decide whether a trust domain is able to protect their resources, additional mechanisms enable participants to verify whether the entities composing the trust domain behaved and currently behave as required. By developing a mechanism to represent Trust Domains and reasons about their properties, we expect to support tool integration and decision-making based on the integration of evidence from disparate sources.